\documentclass[a4paper, 10pt]{article} 
\usepackage{graphicx, amssymb} 
\usepackage[sumlimits]{amsmath}
\usepackage{amsthm}
\input{xy}
\xyoption{all}

\newtheorem{thm}{Theorem}

\newtheorem{fact}[thm]{Fact}
\newtheorem{sub}[thm]{Subroutine}
\newtheorem{alg}[thm]{Algorithm}

\newcommand{\refsec}[1]{Section~\ref{sec:#1}}
\newcommand{\refalg}[1]{Algorithm~\ref{alg:#1}}
\newcommand{\reffig}[1]{Figure~\ref{fig:#1}}
\newcommand{\refeqn}[1]{(\ref{eqn:#1})}
\newcommand{\reftbl}[1]{Table~\ref{tbl:#1}}

\newcommand{\refthm}[1]{Theorem~\ref{thm:#1}}
\newcommand{\refsub}[1]{Subroutine~\ref{sub:#1}}
\newcommand{\reffact}[1]{Fact~\ref{fact:#1}}

\newcommand{\R}{\mathbb R}
\newcommand{\eps}{\varepsilon}
\newcommand{\cA}{{\cal A}}
\newcommand{\cD}{{\cal D}}
\newcommand{\cL}{{\cal L}}
\newcommand{\cP}{{\cal P}}
\newcommand{\cPl}{{\cal P}_{\mathrm{low}}}

\newcommand{\cS}{{\cal S}}
\newcommand{\cM}{{\cal M}}

\newcommand{\If}{I_{\mathrm{free}}}

\newcommand{\smin}{\sigma_{\mathrm{min}}}
\newcommand{\smax}{\sigma_{\mathrm{max}}}
\newcommand{\lmin}{\lambda_{\mathrm{min}}}

\DeclareMathOperator{\spn}{span}
\DeclareMathOperator{\rank}{rank}
\DeclareMathOperator{\diag}{diag}
\DeclareMathOperator{\wsize}{wsize}
\DeclareMathOperator{\proj}{proj}
\DeclareMathOperator{\tr}{Tr}

\newcommand{\pfstart}{\begin{proof}} 
\newcommand{\pfend}{\end{proof}} 

\title{Span-program-based quantum algorithm for the rank problem}
\author{Aleksandrs Belovs\thanks{Faculty of Computing, University of Latvia, Raina bulv. 19, Riga, LV-1586, Latvia, stiboh@gmail.com.}}
\date{}

\begin{document}
\maketitle

\begin{abstract}
Recently, span programs have been shown to be equivalent to quantum query algorithms. It is an open problem whether this equivalence can be utilized in order to come up with new quantum algorithms. We address this problem by providing span programs for some linear algebra problems.

We develop a notion of a high level span program, that abstracts from loading input vectors into a span program. Then we give a high level span program for the rank problem. The last section of the paper deals with reducing a high level span program to an ordinary span program that can be solved using known quantum query algorithms.
\end{abstract}

\section{Introduction}
Span programs, introduced by Karchmer and Wigderson in~\cite{spanFirst}, is a certain way of defining Boolean functions. Initially, they were used over finite fields and applications included, in particular, a log-space analogue of the complexity class inclusion $\mathsf{NP}\subseteq \oplus \mathsf{P}$, and secret sharing schemes.

The realization of their connection to quantum computation is due to the research in quantum algorithms for formulae evaluation. This trend of research was initiated by papers~\cite{formulae} by Farhi {\em et al.} and~\cite{Ambainis} by Ambainis {\em et al.} on computing AND-OR trees. Reichardt and \v Spalek applied span programs over complex numbers while extending the set of allowed gates to all three-variable Boolean functions~\cite{gate3}. The key feature of span programs, that is especially useful for computing formulae, is the ease with that they compose.

Even more, later it has been shown that span programs, measured by a newly defined complexity measure -- witness size, and quantum query algorithms, measured by the number of queries, are essentially equivalent! This was done by Reichardt. At first, up to a logarithmic factor in~\cite{spanBig}. The latter was successfully removed in~\cite{spanOptimal}. The lower bound was proven by showing that the generalized adversary bound~\cite{adversary} by H\o yer {\em et al.} is dual to the witness size of a span program in the sense of semi-definite programming. With the quantum algorithm for evaluating span programs in hand, this proves the equivalence.

Although any Boolean function can be optimally (in the number of queries) evaluated by a span program, it is still an open problem to come up with a good quantum algorithm based on span program evaluation. Until now, the only examples included formula evaluation, like in the already cited~\cite{gate3} or in a more recent~\cite{newFormula}. However, a significant difference of span programs from other models of quantum computation gives a hope they could provide some insights into the construction of efficient quantum algorithms.

In this paper, we embark on extending the family of algorithms based on span programs. It seems natural to start with linear algebra problems, since they are most in the spirit of span programs. We strike a first step in this direction by designing, firstly, an algorithm that is almost a restatement of a span program: decide whether a fixed vector is in the span of given vectors. Secondly, we use this algorithm to solve the rank problem. 

In the rank problem, we are given an $n\times m$-matrix $A$ and an integer $0\le r\le n$. The task is to detect whether $\rank A\ge r$. The most important special case is $r=m=n$. The latter is known as the determinant problem and consists in distinguishing a singular matrix from a non-singular one. Of course, there is nothing special about the zero eigenvalue: if $A$ is a matrix, one can detect whether it has eigenvalue $\lambda$ by considering $A-\lambda I$ as an input to the determinant problem. This is a common building block for many quantum algorithms. For instance, the quantum walk approach by Szegedy~\cite{walk} consists in deciding whether a specific unitary transformation has eigenvalue 1. Also, the optimal quantum algorithm for span programs themselves tests a matrix for having eigenvalue 0~\cite{spanOptimal}.

D\"orn and Thierauf show an $\Omega(n^2)$ lower bound on quantum query complexity of the determinant (and, hence, the rank) problem~\cite{det}. This could be expected, as, in general, a very small change is required for a singular matrix to become non-singular. We, however, consider a promise problem, with a gap between allowed matrices of ranks $<r$ and $\ge r$. 

Let $c_r(A)$ be the quadratic mean of the reciprocals of the $r$ largest singular values of $A$:
$$c_r(A) = \sqrt{\frac1r\left(\frac{1}{\sigma_1^2}+\frac{1}{\sigma_2^2}+\cdots+\frac{1}{\sigma_r^2}\right)}.$$
This value is defined for all matrices of rank at least $r$, and is infinitely large for all other matrices. One of the main contributions of the paper is the following

\begin{alg}
\label{alg:rank}
The rank problem can be solved in $O(\sqrt{r(n-r+1)}LT)$ quantum queries with the promise that any input matrix $A$ has all entries bounded, by absolute value, by 1, and any input matrix, with rank at least $r$, satisfies $c_r(A)\le L$. Here $T$ is the cost of loading an $n\times m$-matrix into a span program.
\end{alg}
Loading a matrix into a span program is a subroutine we deal with in \refsec{loading}. It is similar to Hamiltonian simulation in ordinary quantum algorithms. 

In particular, \refalg{rank} can be used to solve the determinant problem for $n\times n$ matrices in $O(\sqrt{n}c_n(A)T)$ queries, where $A$ is the worst matrix among all allowed non-singular matrices. Value $c_n(A)$, we denote also by $c(A)$, admits an alternative description as 
\begin{equation}
\label{eqn:cAEuclidean}
c(A) = \|A^{-1}\|_E/\sqrt{n},
\end{equation}
where $\|\cdot\|_E$ is the Euclidean norm.

The determinant problem has close affinity to the quantum phase estimation~\cite{phase}. Using the latter it is possible to give a quantum algorithm solving the determinant problem for an $n\times n$ Hermitian matrix $H$ in $O(\sqrt{n}/\lmin(H))$ applications of $e^{iH}$. It is worse than~\refeqn{cAEuclidean}, if $H$ has a broad spectrum. We elaborate more on this subject in \refsec{previous}. It is also possible~\cite{personal} to get a bound similar to~\refeqn{cAEuclidean} by applying a novel variable-time quantum amplitude amplification algorithm by Ambainis~\cite{variableAmplitude}, but our span program is more straightforward. 

The paper is organized as follows. In \refsec{prelim}, we give some basic notions from linear algebra and describe span programs and their complexity measure -- witness size. In \refsec{quantum}, some basic concepts of quantum computation are given. The content of this section is brought into for the purpose of comparison only, and is not used in the main part of the paper. We also describe how a typical quantum algorithm for the determinant problem would look like, if it had appeared in a graduate textbook on quantum computation. We compare it to our span-program-based algorithm in \refsec{comparison}. We conclude that, under a reasonable probability distribution on matrices, our algorithm is asymptotically faster, almost surely. Another contribution of this section is a result on the distribution of the trace of an inverse Wishart matrix. The result is of independent interest and is used in the construction of a span program for the rank problem in \refsec{rank}.

After that, we proceed with high level span programs in \refsec{highLevel}. These are the same span programs as in \refsec{span}, but with the assumption we can query input vectors directly. In \refsec{definition}, we define them, and in \refsec{rank}, describe a high level span program for the rank problem. 

Unfortunately, the novelty of span programs has its downside. We cannot use tools developed in quantum computation, hence, we define some low level machinery for span programs in \refsec{loading}, and this takes a large part of the paper.


\section{Preliminaries}
\label{sec:prelim}
\subsection{Linear Algebra}
\label{sec:linear}
For the basic notions of linear algebra reader may refer to~\cite{horn}. We work with real vector spaces mostly, with an exception for \refsec{quantum} only. We denote the inner product by $\langle x, y\rangle$. By $\|x\|$, we denote the 2-norm of $x$.

If $A$ is an $n\times m$-matrix, we denote by $\|A\|$ its spectral norm $\max_x \|Ax\|/\|x\|$. By $\|A\|_E$, we denote its Euclidean (also known as Frobenius) norm $\sqrt{\tr AA^T}$.

Since $AA^T$ is a positive definite matrix, it has only non-zero eigenvalues. Square roots of the eigenvalues are known as singular values of $A$. Any $n\times m$ matrix admits singular value decomposition $A=U\Sigma V^T$ with $U$ being an $n\times n$ orthogonal (i.e., unitary and real) matrix, $V$ being an $m\times m$ orthogonal matrix, and $\Sigma$ being $n\times m$ matrix with singular values on the ``diagonal'' and all other elements zeroes. The columns of $U$ are called the left singular vectors of $A$, and the columns of $V$ are the right singular vectors of $A$.

The matrix norms defined can be easily described using singular values. Let $\sigma_1,\dots,\sigma_n$ be the singular values of $A$. The spectral norm equals $\smax(A)$, i.e., the maximal singular value of $A$, and the Euclidean norm equals $\sqrt{\sigma_1^2+\cdots+\sigma_n^2}$. These equalities follow directly from the spectral decomposition and the invariance of 2-norm under orthogonal transformations.

\subsection{Span Programs}
\label{sec:span}
In this section, we define span programs following, mostly,~\cite{spanBig}. A span program $\cP$ is a way of computing a Boolean function $\{0,1\}^m\to\{0,1\}$. It is defined by
\begin{itemize}
\item A finite-dimensional inner product space $V=\R^n$. Reichardt {\em et al.} define span programs over $\mathbb{C}$, we find real span programs more convenient. Real span programs are known to be equivalent to the complex ones~\cite[Lemma 4.11]{spanBig};
\item A non-zero {\em target vector} $t\in V$;
\item A set of {\em input vectors} $I\subset V$. The set $I$ is split into the union of the set of {\em free input vectors} $\If$ and the collection of sets $\{I_{j,b}\}$ with $j=1,\dots,m$ and $b=0,1$: $I=\If\cup\bigcup_{j,b} I_{j,b}$. The input vectors of $I_{j,b}$ are {\em labeled} by the tuple of the $j$-th input variable $x_j$ and its possible value $b$.
\end{itemize}

For each input $x=(x_j)\in\{0,1\}^m$, define the set of {\em available} input vector as $I(x)=\If\cup\bigcup_{j=1}^m I_{j,x_j}$. Its complement $I\setminus I(x)$ is called the set of {\em false} input vectors. We say that $\cP$ evaluates to 1 on input $x$, iff $t\in\spn(I(x))$. In this way, span programs define total Boolean functions. One can define a span programs for a partial Boolean function as well, by ignoring the output of the program on the complement of the domain.

A useful notion of complexity for a span program is that of {\em witness size}. Assume, up to the end of the section, a span program $\cP$ calculates a partial Boolean function $f:\cD\to\{0,1\}$ with $\cD\subseteq \{0,1\}^m$. Let $A$ and $A(x)$ be matrices having $I$ and $I(x)$ as their columns, respectively.

If $\cP$ evaluates to 1 on input $x\in\cD$, a {\em witness} for this input is any vector $w\in \R^{|I(x)|}$ such that $A(x)w = t$. The size of $w$ is defined as its norm squared $\|w\|^2$. 

If, on contrary, $f(x)=0$ then a witness for this input is any vector $w'\in V$ such that $\langle w',t\rangle=1$ and that is orthogonal to all vectors from $I(x)$. Since $t\notin\spn(I(x))$, such a vector exists. The size of $w'$ is defined as $\|A^T w'\|^2$. Note that this equals the sum of squares of inner products of $w'$ with all false input vectors.

The witness size $\wsize(\cP, x)$ of span program $\cP$ on input $x$ is defined as the minimal size among all witnesses for $x$ in $\cP$. We also use notation
$$\wsize_b(\cP,\cD) = \max_{x\in \cD: f(x)=b} \wsize(\cP,x).$$
The witness size of $\cP$ is defined as
$$\wsize(\cP,\cD) = \sqrt{\wsize_0(\cP,\cD)\wsize_1(\cP,\cD)}.$$
This is not a standard definition, but it appears as equation (2.8) in~\cite{spanBig}.

The following important theorem is a combination of results from~\cite{spanOptimal} and~\cite{spanBig} and it shows why span programs are important for quantum computation:
\begin{thm}
\label{thm:span}
For any partial Boolean function $f\colon \{0,1\}^n\supset \cD\to \{0,1\}$ and for any span program $\cP$ computing $f$, there exists a 2-sided bounded error quantum algorithm calculating $f$ in $O(\wsize(\cP, \cD))$ queries.
\end{thm}

Thus, a search for a good quantum query algorithm is essentially equivalent to a search for a span program with small witness size.

\section{Previous Results}
\label{sec:previous}
The main point of this section is to interpret \refalg{rank} in the context of known quantum algorithms. In order to simplify the comparison, we limit ourselves to the determinant problem, as the most important special case.

In \refsec{quantum}, we give a short exposition of quantum algorithms we find relevant to the determinant problem. The content of this section is not used in the proofs of the main results. However, we utilize some results to prove some lower bounds in \refsec{lower}.

Also, as a quantum algorithm for the determinant problem doesn't seem to appear in an explicit form in the literature, we give our variant, \refalg{convenient}, based on standard quantum subroutines. In \refsec{comparison}, we compare the performance of Algorithms~\ref{alg:rank} and~\ref{alg:convenient}. In order to do this, we use Gaussian matrices, and prove a result on the distribution of the trace of an inverse Wishart matrix. The content of this section is used later in the proof of \refalg{rank} in \refsec{rank}.

\subsection{Quantum Query Algorithms}
\label{sec:quantum}
For the basic concepts of quantum algorithms, a reader may refer to~\cite{chuang}. Query algorithms measure the complexity of a problem by the number of queries to the input the best algorithm should make. Clearly, it provides a lower bound on the time complexity. For many algorithms, query complexity can be analyzed easier than time complexity. For the definition of query complexity and its basic properties, a good reference is~\cite{survey}.

One of the basic quantum algorithms is Grover search~\cite{grover}. It is capable of computing the OR function in $O(\sqrt{n})$ quantum queries. Its optimality was one of the first quantum lower bound results. The Grover algorithm is optimal even for the {\em unique search problem}, when it is promised the input string contains no more than 1 element set to `1'.

\begin{thm}[\cite{groveroptimal}]
\label{thm:grover}
Any quantum algorithm discriminating zero string from any string, containing exactly one `1', requires $\Omega(\sqrt{n})$ quantum queries.
\end{thm}

One extension of Grover search is quantum amplitude amplification~\cite{amplitude}. Assume $\cA$ is a quantum algorithm without measurements and let one of its registers, $x$, represent the ``goodness'' of the output: the output is considered {\em good} iff $x=1$. Assume $a$ is the probability of obtaining a good output when the final output of $\cA$ is measured in the standard basis. Then, quantum amplitude amplification allows one to boost this probability up to $\Omega(1)$ in $O(1/\sqrt{a})$ applications of $\cA$. 

Two quantum subroutines are important for us. The first one is Hamiltonian simulation. The problem is to (approximately) implement the unitary $e^{iH}$ where $H$ is a Hermitian matrix encoded with input variables. Recently, Childs and Kothari came up with an improved algorithm~\cite{childs} for the Hamiltonian simulation problem running in time $(d+\log^*N)d^2\|H\|(d\|H\|/\delta)^{o(1)}$ where $d$ is the maximal number of non-zero entries in any row of $H$ and $\delta$ is precision of the algorithm. The algorithm, additionally, assumes matrix $H$ is efficiently row-computable, i.e., for any row index $i$, one can get a $d$-list containing indices of all non-zero elements in the $i$-th row of $H$.

Another important subroutine is quantum phase estimation~\cite{phase}. Given a unitary $U$ and its eigenvector $\psi$, the algorithms produces $\phi$ such that $e^{i\phi}$ is the eigenvalue corresponding to $\psi$. If $\eps$ is error probability of the algorithm and $\delta$ is precision (i.e., an answer within $\pm \delta$ interval around the true value is considered as correct), the algorithm can be implemented using $O(\frac1\eps \log\frac1\delta)$ controlled applications of $U$~\cite{phaseNew}. The algorithm can be extended to all $\psi$'s, that are not necessary eigenvectors of $U$, by linearity.

Since it has not been done explicitly before, we compose these quantum subroutines into an algorithm for solving the determinant problem. Recall, in the determinant problem, we are given an $n\times n$ matrix $A$ and we should decide whether it has full rank. In \refsec{comparison}, we compare this algorithm to our span-program-based \refalg{rank}.

\begin{alg}
\label{alg:convenient}
The determinant problem can be solved in $\tilde O(\sqrt{n}LT)$ quantum queries with the promise that any input matrix $A$ satisfies $\|A\|\le 1$ and any non-singular input matrix $A$ satisfies $\smin(A)\ge 1/L$. Here, $T$ is the number of queries needed to simulate $A$ (see further in the proof).
\end{alg}

\pfstart[Proof sketch]
Let $\{\sigma_i\}$ be the singular values of $A$. Since $A$ is not Hermitian, it cannot be used directly in the Hamiltonian simulation subroutine. The solution is to replace $A$ by
$$H = \begin{pmatrix}0 & A \\ A^* & 0\end{pmatrix},$$
where $A^*$ is the transpose complex conjugate of $A$. The matrix $H$ is Hermitian and has $\{\pm\sigma_i\}$ as eigenvalues. Parameter $T$ in the statement of the algorithm describes the cost of simulating $H$. Note also, that if one wishes to apply sparse Hamiltonian simulation, it is enough for $H$ to be efficiently row-computable, but this means that $A$ should be both efficiently row- and column-computable.

By the norm bound on $A$, the unitary $e^{iH}$ has eigenvalue 1, iff $A$ is singular. The next step is to apply phase estimation on $e^{iH}$ with a random vector as $\psi$. In order to distinguish between zero and non-zero eigenvalues of $H$, the precision of the phase estimation should be at least $1/L$.

Say the output of the phase estimation is good, if the phase $\phi=0$. If $A$ has rank $n-1$, the probability of obtaining a good output when measuring the final state is $1/n$. If $A$ has full rank, it can be made much smaller.

Using quantum amplitude amplification, we can boost the probability $1/n$ up to $\Omega(1)$ in $O(\sqrt{n})$ applications of the phase estimation. This makes the total $\tilde O(\sqrt{n}L)$ applications of $e^{iH}$.
\pfend

\subsection{Comparison of Algorithms}
\label{sec:comparison}
We have seen two algorithms for the determinant problem: Algorithms~\ref{alg:convenient} and~\ref{alg:rank}. The first one requires $O(\sqrt{n}/\smin)$ applications of $e^{iH}$, whereas the second one uses the number of queries equal to $O(\sqrt{n}c(A))$ times the complexity of loading $A$ into the span program. We stick to our view that loading a matrix into a span program is an operation of the same flavor as simulating a Hamiltonian, hence, we cancel out the corresponding factors in both of the complexity estimations.

Under this assumption, the complexity comparison boils down to comparing $1/\smin(A)$ and $c(A)$. It is easy to see, the latter is smaller than the former, but is is not clear whether the improvement is significant. Of course, it depends on the type of a problem the algorithm is used to solve and the matrices appearing therein. If the problem is not set, it is natural to compare performance on random instances. In this section, we consider a natural probability distribution on matrices and show that, under this distribution, $c(A)$ is asymptotically smaller than $1/\smin(A)$ with the ratio being $\Theta(\sqrt{n})$.

A {\em Gaussian matrix } $G(n, m)$ is an $n\times m$ random matrix with entries independently drawn from the standard normal distribution $N(0,1)$. This is one of the natural probability distributions on matrices. We prove that if $A$ is taken from $G(n,n)$ then, almost surely, $c(A)$ is asymptotically smaller than $1/\smin(A)$. 

Later, we will need 2 well-known facts about normal distributions we state here:

\begin{fact}
\label{fact:linear}
A linear combination of independent normal distributions $c_1N(0,\sigma_1^2)+c_2N(0,\sigma_2^2)$ is the normal distribution $N(0,c_1^2\sigma_1^2+c_2^2\sigma_2^2)$.
\end{fact}

\begin{fact}
\label{fact:orthogonal}
Gaussian matrix distribution $G(n,m)$ is invariant under multiplication by orthogonal matrices both from the left and from the right.
\end{fact}

If $A\sim G(n,m)$ then the distribution of $W=AA^T$ is known as the {\em Wishart distribution} $W(n,m)$. It is a probability distribution on symmetric $n\times n$ matrices. Parameter $m$ is known as the {\em number of degrees of freedom}. The Wishart distribution is a generalization of the chi-square distribution, since $W(1,m)$ coincides with $\chi_m^2$. The Wishart distribution is important for us because $\smin(A)^2 = \lmin(W)$ and $c(A)^2 = \frac1n\tr(W^{-1})$. The following is known about the Wishart distribution. For Theorems~\ref{thm:inverseWishart} and~\ref{thm:bartlett}, refer, e.g., to Section 3.2 of~\cite{muirhead}.

\begin{thm}
\label{thm:inverseWishart}
If $m>n+1$ then $E[W^{-1}] = \frac{1}{m-n-1}I_n$ when $W\sim W(n,m)$.
\end{thm}

\begin{thm}[Edelman~\cite{edelman}]
\label{thm:minimalWishart}
If $W\sim W(n,n)$, then as $n\to\infty$, $n\lmin(W)$ converges in distribution to a random variable with the probability density function
\begin{equation}
\label{eqn:minimalWishart}
\frac{1+\sqrt{x}}{2\sqrt{x}}e^{-(x/2+\sqrt{x})}.
\end{equation}
\end{thm}

The Cholesky decomposition of $W(n,m)$ is also well-understood:
\begin{thm}[Bartlett decomposition]
\label{thm:bartlett}
A Wishart matrix $W(n,m)$ equals $TT^T$ where
\begin{equation}
\label{eqn:bartlett}
T = \begin{pmatrix}
\sqrt{t_1} & 0 & 0 & \cdots & 0 \\
t_{21} & \sqrt{t_2}     & 0 & \cdots & 0 \\
t_{31} & t_{32} &  \sqrt{t_3}      & \cdots & 0 \\
\vdots & \vdots & \vdots & \ddots & \vdots \\
t_{n1} & t_{n2} & t_{n3} & \cdots & \sqrt{t_n}
\end{pmatrix}\end{equation}
with $t_i\sim \chi_{m-i+1}^2$ and $t_{ij}\sim N(0,1)$ independently.
\end{thm}

In particular, \refthm{minimalWishart} means that $\smin(A)=O(1/\sqrt{n})$ almost surely, and this solves the complexity estimation for the quantum phase estimation algorithm. The following theorem gives an estimate on the distribution of $\tr(W^{-1})$ and shows that $c(A)$, almost surely, remains bounded by a constant. Interestingly, we were not able to find a similar statement proven prior to our paper. 

\begin{thm}
\label{thm:traceWishart}
For any $\eps>0$, there exists $\delta$ such that, for any $n$, $Pr[c(A)>\delta]\le \eps$ when $A\sim G(n,n)$.
\end{thm}

\pfstart
Let $W=AA^T$. One way to approach the theorem would be to get an expectation for the trace of $W^{-1}$ and then apply Markov inequality. But unfortunately, we cannot apply \refthm{inverseWishart} directly, and there is a good reason for that: the expectation doesn't exist. It's mostly because the expectation of the inverse of~\refeqn{minimalWishart} doesn't exist. We solve this complication, broadly speaking, by using \refthm{minimalWishart} for the smallest two singular values of $A$ and applying \refthm{inverseWishart} for the rest of the matrix.

We utilize the Bartlett decomposition. Thus, let $T$ be as in~\refeqn{bartlett} with $t_i\sim\chi_{n-i+1}^2$, so that $W = TT^T$. Clearly, $A$ and $T$ have the same singular values, and the same holds for $A^{-1}$ and $T^{-1}$. In particular, because of~\refeqn{cAEuclidean}, we have
\begin{equation}
\label{eqn:ncA}
n\, c(A)^2=\|T^{-1}\|_E^2.
\end{equation}
To estimate the latter, represent $T$ in the block-diagonal form:
$$T = \begin{pmatrix}
T_{11} & 0\\
T_{21} & T_{22}
\end{pmatrix},$$
where $T_{11}$ and $T_{22}$ are square matrices of dimensions $n-2$ and 2, respectively. Then calculate the inverse
\begin{equation}
\label{eqn:Tinv}
T^{-1} = 
\begin{pmatrix}
\tilde T_{11} & 0 \\
\tilde T_{21} & \tilde T_{22}
\end{pmatrix}=
\begin{pmatrix}
T_{11}^{-1} & 0  \\
-T_{22}^{-1}T_{21}^{}T_{11}^{-1} & T_{22}^{-1}
\end{pmatrix}.
\end{equation}

Denote $W_{11}=T_{11}T_{11}^T$ and notice it is the Bartlett decomposition of $W(n-2,n)$. Hence, by \refthm{inverseWishart} and the linearity of expectation, 
\begin{equation}
\label{eqn:Evklid1}
E[\|T_{11}^{-1}\|_E^2] = E[\tr W_{11}^{-1}] = n-2.
\end{equation}

A norm of a row of a matrix is bounded by its largest singular value (i.e., its spectral norm). Applying this to the last two rows of $T^{-1}$, one gets
\begin{equation}
\label{eqn:Evklid2}
\|\tilde T_{21}\|_E^2 + \|\tilde T_{22}\|_E^2 \le 2(\smax(T^{-1}))^2 = 2 /\lmin(W).
\end{equation}

Applying Markov inequality to~\refeqn{Evklid1}, one obtains a constant $\delta_1$ such that 
\begin{equation}
\label{eqn:ocenka1}
Pr[\|T_{11}^{-1}\|_E^2>\delta_1n]\le \eps/2.
\end{equation}
On the other hand, by \refthm{minimalWishart}, there exists $\delta_2>0$ such that 
\begin{equation}
\label{eqn:ocenka2}
Pr[\lmin(W) < \delta_2/n]\le\eps/2.
\end{equation}
Plugging~\refeqn{ocenka2} into~\refeqn{Evklid2} and combining with~\refeqn{ocenka1}, in the sight of~\refeqn{ncA} and~\refeqn{Tinv}, we get that 
$$Pr[c(A)>\delta_1+2/\delta_2]\le \eps,$$
independently on $n$.
\pfend

\section{High Level Span Programs}
\label{sec:highLevel}
In this section, we define high level span programs. The idea of this definition is in the assumption we can query input vectors in a span program directly. Recall that in an ordinary span program (that we also call {\em actual} span program, or a {\em low level} span program), the set of possible input vectors is fixed, and these are the input Boolean variables that specify which of them become available for the span program. For simplicity, we consider real span programs only, complex span programs can be simulated by the real ones using standard techniques.

Of course, querying vectors is not a commonly admitted operation. In \refsec{loading}, we give a reduction of a high level span program to an ordinary span program. We call this operation {\em matrix loading}.

\subsection{Definition}
\label{sec:definition}
The specification of a high level span program $\cP$ consists of
\begin{itemize}
\item two positive integers $n$ and $m$. The first one specifies the dimension of the vector space, and the second one specifies the number of input vectors. We represent the input vector as columns of an $n\times m$-matrix $A$. The $j$-th input vector is denoted by $a_j$;
\item a fixed non-zero target vector $t\in\R^n$;
\item a set of valid input matrices $\cD$;
\item a subspace of free input vectors $F$. It is an optional ingredient of a span program, but it is often easier to describe a span program using free input vectors. Usually, $F$ is given as a span of a finite set of free input vectors.
\end{itemize}

The span program distinguishes whether affine subspace $t+F$ intersects the linear span of columns of $A$ (we denote the latter subspace by $\spn(A)$), or not. Clearly, it is a hard problem to decide whether a vector is inside a subspace, or the subspace is perturbed slightly to not contain it. Because of this problem, and inspired by the definitions from \refsec{span}, we define the complexity measure of a high level span program called {\em witness size}.

If $A\in \cD$ is such that $t+F$ intersects $\spn A$, we define its {\em witness} as any vector $w\in\R^m$ such that $Aw\in t+F$. If, on contrary, $A$ is such that $t+F$ and $\spn A$ do not intersect, we define its witness as a vector $w'\in\R^n$ with the property $\langle w', t\rangle=1$, $w'\perp \spn A$ and $w'\perp F$. We call the former case a {\em positive} one, and the latter case a {\em negative} one.

The {\em witness size} of a valid input $A$ is defined as 
$$\wsize(P, A) = \min\{\|w\|^2 \mid \mbox{$w$ is a witness for $A$ in program $P$}\}.$$
The positive- and negative-input witness sizes are defined as 
$$\wsize_1(\cP) = \max_{\substack{A\in \cD\\ (t+F)\cap\, \spn A\ne\emptyset }} \wsize(\cP,A)\qquad\mbox{and}\qquad \wsize_0(\cP) = \max_{\substack{A\in \cD\\ (t+F)\cap\, \spn A=\emptyset}} \wsize(\cP,A).$$
And finally, the witness size of $\cP$ is defined as the geometric mean of its positive- and negative-input witness sizes:
$$\wsize(\cP) = \sqrt{\wsize_0(\cP)\wsize_1(\cP)}.$$

It is easy to see that the result of the program (but not the witness size!) remains unchanged if we rescale the input vectors. So, for convenience, we will latter assume that all entries of the input matrix are from interval $[-1,1]$.

The following theorem is a restatement of Subroutines~\ref{sub:loading1}, \ref{sub:loading2} and \ref{sub:loading3}.
\begin{thm}
\label{thm:highLevel}
Any high level span program $\cP$ can be solved by a quantum query algorithm of complexity $O(\wsize(P)L)$ where $L$ is complexity of loading matrix $A$ into the span program.
\end{thm}

\subsection{Span Program for Rank Problem}
\label{sec:rank}
In this section, we give a high level span program for the rank problem, \refalg{rank}. Recall that in the rank problem, we are given an $n\times m$-matrix $A$ and an integer $0\le r\le n$. The task is to detect whether $\rank A\ge r$. In general, one may assume $n\le m$, because, otherwise, the matrix can be transposed reducing the complexity of the algorithm. Also, recall the definition of complexity measure $c_r(A)$ from the introduction:
$$c_r(A) = \sqrt{\frac1r\left(\frac{1}{\sigma_1^2}+\frac{1}{\sigma_2^2}+\cdots+\frac{1}{\sigma_r^2}\right)}.$$
where $\sigma_1\ge \sigma_2\ge \cdots \ge \sigma_r>0$ are the $r$ largest singular values of $A$. We repeat the statement of the algorithm here:

\newcounter{asdf}
\setcounter{asdf}{\value{thm}}
\setcounter{thm}{0}

\begin{alg}
The rank problem can be solved in $O(\sqrt{r(n-r+1)}LT)$ quantum queries with the promise that any input matrix $A$ has all entries bounded, by absolute value, by 1, and any input matrix, with rank at least $r$, satisfies $c_r(A)\le L$. Here $T$ is the cost of loading an $n\times m$-matrix into a span program.
\end{alg}

\setcounter{thm}{\value{asdf}}

\pfstart
Denote $s=n-r$. Given vectors $a_1,\dots,a_m$, we add $s$ free input vectors $v_1,v_2,\dots,v_{s}$ and check if a random vector $t$ is contained in the span of $\{a_1,\dots,a_m,v_1,\dots,v_s\}$. The idea is that if the rank of $A$ is less than $r$, adding $s$ vectors to it won't make it to have full rank. On contrary, if $\rank A \ge r$, vectors $a_1,\dots,a_m,v_1,\dots,v_s$ span the whole space with probability 1. We generate $v_1,\dots,v_s$ and $t=(t_i)$ by letting their entries be independent standard Gaussians. 

At first, we consider the case $\rank A\ge r$ and estimate the witness size. For simplicity, we assume the rank of $A$ is exactly $r$. Otherwise, the witness size can only decrease. Let $\sigma_1,\dots,\sigma_r$ be the largest singular values of $A$. Let $V=(v_{ij})$ be the matrix with $v_j$'s as columns. Thus, $V\sim G(n,s)$. Because of \reffact{orthogonal}, we may assume the elements of the standard basis $e_1,\dots,e_r$ are equal to the left singular vectors of $A$. Hence, $A=\Sigma_A O_A$ where $\Sigma_A$ is $n\times m$ matrix with $\sigma_1,\dots,\sigma_r$ on the ``diagonal'' (and all other elements zeroes) and $O_A$ is an $m\times m$ orthogonal matrix. Let $\tilde V$ be the $s\times s$ matrix formed by the last $s$ rows of $V$. Similarly, we denote by $\tilde t$ the last $s$ elements of $t$.

We proceed as follows. At first, we find $x\in\R^s$ such that $\tilde Vx = \tilde t$. Then $t'=t-Vx$ is in the span of $A$. Thus, we search for $w$ such that $Aw = t'$. The witness size of $A$ is at most $\|w\|^2$.

According to \refthm{traceWishart}, we choose $\delta$, independently on $n$ and $s$, such that 
\begin{equation}
\label{eqn:ctildeV}
Pr[c(\tilde V)\le \delta]\ge 11/12,
\end{equation}
and condition that this is the case, i.e., $c(\tilde V)$ is bounded by $\delta$. Fix $\tilde V$ and denote its singular values by $\sigma_{r+1},\dots,\sigma_n$. Again, we may assume $e_{r+1},\dots,e_n$ are the left singular vectors of $\tilde V$. Hence $\tilde V=\Sigma_VO_V$ with $\Sigma_V = \diag\{\sigma_{r+1},\dots,\sigma_n\}$ and $O_V$ orthogonal. Under the assumption on $c(\tilde V)$, vector $x$, with the property $\tilde Vx=\tilde t$, satisfies
$$E\left[\|x\|^2\right] = E\left[\|O_Vx\|^2\right] = E\left[\sum_{i=r+1}^n t_i^2/\sigma_i^2 \right] = s\, c(\tilde V)^2 = O(s),$$
since, for the standard Gaussian $t_i$, one has $E[t_i^2]=1$. 

Now fix $x=(x_j)$. The first $r$ coordinates of $t$ and $v_j$'s are independent on $x$. Hence, each of the first $r$ coordinates of $t'=(t_i')=t-Vx$, as a linear combination of independent normal distributions, by \reffact{linear}, has distribution $N(0,1+\|x\|^2)$. In particular, unfixing $x$, we have
$$E[t_i'^2] = 1 + E[\|x\|^2] = O(1+s).$$
And, finally, for $w$, satisfying $Aw=t'$, by the linearity of expectation, we have:
$$E\left[\|w\|^2\right] = \sum_{i=1}^r \frac{1}{\sigma_i^2} E[t_i'^2] = O\left((1+s)r\, c_r(A)^2\right).$$

By Markov inequality, $Pr\left[\|w\|^2 > 12 E[\|w\|^2]\right]<1/12$. Combining this with~\refeqn{ctildeV} and the assumption on $c_r(A)$, we have that $\|w\|^2 = O((n-r+1)r L^2)$ with probability $5/6$.

Now assume $A$ has rank less than $r$. In this case, $A$, together with $v_i$'s, does not span the whole space. We can assume $e_1$ is orthogonal to the span of $\{a_1,\dots,a_m,v_1,\dots,v_s\}$. Then, the witness of $A$, inside the one-dimensional subspace spanned by $e_1$, has size $1/t_1^2$. It is $O(1)$ with probability $5/6$.

In order to solve the rank problem, we execute the span program with the bound $O((n-r+1)r L^2)$ on $\wsize_1$ and $O(1)$ on $\wsize_0$. We require error probability of the span program solving algorithm be at most $1/6$. The total error probability is less than $1/3$, and the query complexity, by \refthm{highLevel}, is $O(\sqrt{r(n-r+1)}LT)$.
\pfend

The complexity of \refalg{rank} is optimal, at least for the $O(\sqrt{r(n-r+1)})$ factor. We show this using the {\em Hamming-weight threshold} function $T_r^n: \{0,1\}^n\to \{0,1\}$ defined in~\cite{spanBig} by
$$T_r^n(x) = \begin{cases}1,&\text{if $|x|\ge r$;}\\0,&\text{otherwise;} \end{cases}$$
where $|x|$ is the Hamming weight of $x=(x_i)$. It is known that any quantum algorithm for $T_r^n$ requires $\Omega(\sqrt{r(n-r+1)})$ queries and there exists a (low level) span program for this function with witness size at most $\sqrt{r(n-r+1)}$~\cite{spanBig}.

The threshold function can be reduced to the rank problem by considering the matrix $A_x = \diag(x)$. It is clear that $\rank A_x \ge r$, if and only if $T_r^n(x) = 1$. For each $A_x$ of rank at least $r$, $c_r(A_x)=1$. The complexity of loading $A_x$ into the span program is $O(\log n)$, as it is shown in \refsub{loading3} further in the text.

Hence, the $O(\sqrt{r(n-r+1)})$ factor in the statement of \refalg{rank} is tight up to a logarithmic factor. Also, \refalg{rank} itself can be considered as a generalization of the span program for the threshold function.

\section{Matrix Loading Subroutines}
\label{sec:loading}
In this section, we develop machinery to deal with high level span programs. We give two subroutines: vector loading and demultiplexor, and then compose them to reduce a high level span program to an actual span program. Span program composition has been developed before, but one difference of our subroutines is that they result in vectors, not in Boolean variables as it was in~\cite{spanBig}.

We give, in total, three different variants of realizing a high level span program by an actual one, each with different complexity and different assumptions on matrix $A$ and its accessibility via queries.

One more convention should be made. In all these subroutines, elements being queried are integers and real numbers. Since span programs have been developed for Boolean variables only, a representation of numbers using Boolean variables should be chosen. So, from now forth, it is always assumed that an integer $c$ in range $[0,n-1]$ is represented by $k=\lceil\log n\rceil$ Boolean variables $c_0,c_1,\dots, c_{k-1}$, so that 

\begin{equation}
\label{eqn:integer}
c=\sum_{i=0}^{k-1} c_i2^i.
\end{equation}

All real numbers are assumed to be in range $[-1,1]$. Any such number $x$ is defined using a number of Boolean variables $x_0,x_1,\dots,x_k$, where $k$ is some predefined precision parameter, and it holds that 
\begin{equation}
\label{eqn:real}
x = \sum_{i=0}^k x_i2^{-i}-1.
\end{equation}

\subsection{Vector producing subroutines}
\begin{figure}[htb]
$$\xymatrix@R=3pt{
\cP = \cP_\ell  & & \cP_{\ell-1} \ar[dl] && \cP_2\ar[dl] && \cP_1\ar[dl] & & \\
& \bullet\ar[ul] && \cdots\ar[ul] && \bullet\ar[ul] && \bullet\ar[ul] \\
\\
& \save [].[0,7]*[F.]\frm{} \restore 
\cS_{\ell-1}\ar[uu] & &\cdots & &\cS_{1}\ar[uu] & & \cS_{0}\ar[uu] & \cP_0\ar[uul]
}$$
\caption{Composing vector producing subroutines into high level span programs. Bullets represent composition operations, $\cP_\ell$ is the final high level span program, and boxed are the components of the actual span program $\cPl$.} \label{fig:compose}
\end{figure}
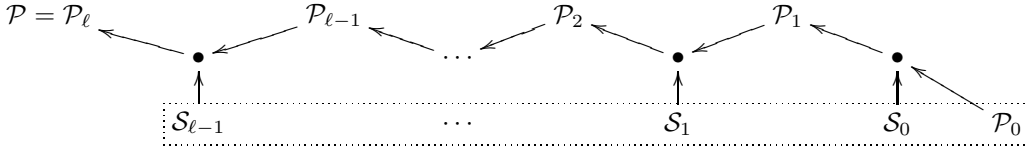

Our goal is to start with a high level span program $\cP$, and end up with a low level span program $\cPl$ calculating the same function. Once this is done, one can apply \refthm{span} to the latter one. Some input vectors of the high level program can be free. In this case, we add them to the set $\If$ of free vectors of $\cPl$. If an input vector is actually queried, we have to implement this query using the Boolean variables representing the vector. We do this in a number of steps, where each step involves composing a {\em vector producing subroutine} into a high level span program. Each composition transforms a high level span program into a high level span program, refer to \reffig{compose}. In this Figure, $\ell$ compositions are performed: each $\cP_i$, for $i=0,\dots,\ell-1$, is composed with $\cS_i$ to give $\cP_{i+1}$. The initial high level span program $\cP_0$ contains only free input vectors. The final result of the composition is $\cP_\ell=\cP$. The actual span program $\cPl$ is the union of $\cP_0$ and all $\cS_i$'s.

Consider the operation of composing $\cS$ into $\cP_1$ in order to get $\cP_2$. The vector space of $\cPl$ contains three vector spaces:
\begin{itemize}
\item the vector space $V_2$ of $\cP_2$. It contains all input vectors of $\cP_2$;
\item the input space $V_1$ of $\cS$. It contains additional dimensions that input vectors of $\cP_1$ are allowed to use, i.e., the composition operation can shrink the vector space of the high level span program. Some reference basis of $V_1$ is fixed. When specifying the composition, a position of $V_1$, relatively to $V_2$, gets chosen. The vector space of $\cP_1$ becomes $V_1 + V_2$;
\item the working space $V_3$ of $\cS$. It is orthogonal to $V_1+V_2$. The input vectors of $\cPl$, labeled by the input variables used in $\cS$, lie inside $(V_1+ V_2)\oplus V_3$.
\end{itemize}

Thus, the vector space of the actual span program $\cPl$ is the direct sum of the vector space of $\cP_0$ and the working spaces of all $\cS_i$'s. All these subspaces are pairwise orthogonal. There are some additional components of the subroutine to be specified:
\begin{itemize}
\item a number of pivots, that are vectors identified with elements of $V_2$ during composition. The subroutine does not affect the components of input vectors of $\cP_1$ outside the span of the pivots and the input space;
\item input variables. These are Boolean variables labeling input vectors used in $\cal S$.
\end{itemize}

The action of the subroutine depends on whether the final outcome of $\cP$ is positive or negative. If it is positive, input vectors of $\cal S$ are used to transform input vectors of $\cP_1$ into input vectors of $\cP_2$. The resulting input vectors should be contained in $V_2$. Also, the witness $w$ of $\cP_2$ is transformed into a witness $\tilde w$ of $\cP_1$.

If the final result is negative, the witness $w'$ of $\cP_2$ gets extended onto $(V_1+V_2)\oplus V_3$ so that it is still a negative witness, i.e., it is orthogonal to all input vectors of $\cP_1$ and all available input vectors of $\cS$. If $V_1$ is orthogonal to $V_2$, it should be always doable (i.e., the subroutine is responsible for that). If $V_1$ intersects $V_2$, it should be additionally assured that this extension is possible.

The total witness size of $\cPl$ breaks naturally into individual {\em costs} of $\cS_i$'s.
That is, the cost of $\cal S$ is the contribution to the witness size from the input vectors of $\cal S$. It depends on the witness of $\cP_2$. 

For the rest of the section, we use notations $w=(w_j)$, $w'=(w'_i)$, $\tilde w = (\tilde w_j)$ and $\tilde w' = (\tilde w'_i)$ for positive and negative witnesses of $\cP_2$ and $\cP_1$, respectively. The last notation $\tilde w'$ will be used also for the negative witness of the whole program, i.e., including the working space of $\cS$.

\subsection{Dense matrices}
We start with a simple vector loading subroutine:

\begin{sub}[Vector Loading]
\label{sub:loading}
There exists a subroutine producing a vector whose coordinates in the pivots of the subroutine are specified by the input real numbers. The specification of the subroutine is in \reftbl{loading}.
\end{sub}

\begin{table}[htb]
\begin{center}
\begin{tabular}{lp{10cm}}
\hline
Pivots:& a system of $n$ vectors $e_1,e_2,\dots, e_n$.\\
Input variables:& $n$ real numbers $\{x_i\}$, $i=1,\dots,n$.\\
Input space:& none.\\
Result:& additional input vector $a_j=\sum_{i=1}^n x_ie_i$, let its index in $\cP_2$ be $j$. \\
Positive witness:& the $j$-th coordinate of $w$ gets removed.\\
Positive cost:& $O(nw_j^2)$.\\
Negative witness:& does not change.\\
Negative cost:& $O(\sum_{i=1}^n \langle w', e_i\rangle^2)$.\\
\hline
\end{tabular}
\end{center}
\caption{Specification of vector loading subroutine} \label{tbl:loading}
\end{table}

\pfstart
Recall that each $x_i$ is given by the sequence of input Boolean variables $\{x_{i,a}\}$ that specify $x_i$ like in~\refeqn{real}. Let the basis of the working space be $f_{i,a}$, with the same range for $i$ and $a$ as in $\{x_{i,a}\}$. For each input digit $x_{i,a}$ and each its possible value $b\in\{0,1\}$, define input vector
$$v_{i,a,b} = b 2^{-a/2} e_i - f_{i,a}.$$
Additionally, define a free input vector
$$v = \sum_{i=1}^n\sum_{a=0}^k 2^{-a/2}f_{i,a} - \sum_{i=1}^n e_i.$$

For the positive case, we show how to construct the resulting vector $a_j$ from the available input vectors. We do this by letting 
\begin{equation}
\label{eqn:koefficienty}
\begin{tabular}{rcl}
the coefficient of & $v_{i,a,x_{i,a}}$ & be equal to $2^{-a/2}$; \\
and the coefficient of & $v$ & be equal to $1$.
\end{tabular}
\end{equation}

Indeed, this linear combination equals
$$
\sum_{i=1}^n\sum_{a=0}^k 2^{-a/2}\left(x_{i,a}2^{-a/2} e_i - f_{i,a}\right) + \left(\sum_{i=1}^n\sum_{a=0}^k 2^{-a/2}f_{i,a} - \sum_{i=1}^n e_i\right) 
=\sum_{i=1}^n \left(\sum_{a=0}^k x_{i,a}2^{-a} - 1\right) e_i =a_j.
$$
In order to get $w_j a_j$ for $\cP_2$, all coefficients of~\refeqn{koefficienty} should be multiplied by $w_j$, that gives the cost of the subroutine equal to
$$\sum_{i=1}^n\sum_{a=0}^k w_j^2 2^{-a} \le 2nw_j^2.$$

For the negative case, we extend the witness $w'$ to the witness $\tilde w'$ of the resulting program by letting
$$\langle \tilde w', f_{i,a}\rangle = x_{i,a}2^{-a/2}\langle w', e_i\rangle.$$
It is trivially orthogonal to all $v_{i,a,x_{i,a}}$'s. It is also orthogonal to $v$, since
$$\left< \tilde w', \sum_{i=1}^n\sum_{a=0}^k 2^{-a/2}f_{i,a} - \sum_{i=1}^n e_i \right> = \sum_{i=1}^n\sum_{a=0}^k 2^{-a}x_{i,a}\langle w', e_i\rangle - \sum_{i=1}^n \langle w', e_i\rangle = \sum_{i=1}^n x_i \langle w', e_i\rangle = \langle w', a_j\rangle = 0.$$

The inner product of $\tilde w'$ with the false input vector, labeled by $x_{i,a}$, equals
$$\langle \tilde w', (1-x_{i,a}) 2^{-a/2} e_i - f_{i,a}\rangle = \langle w', e_i\rangle (1-x_{i,a}) 2^{-a/2} - x_{i,a} 2^{-a/2}\langle w', e_i\rangle = (1-2x_{i,a}) 2^{-a/2}\langle w', e_i\rangle.$$
And the total contribution to the witness size is no more than $2 \sum_{i=1}^n \langle w', e_i\rangle^2.$
\pfend

Applying this procedure $m$ times gives the following theorem that is applicable to any $n\times m$-matrix if we assume the matrix is given as a table of $nm$ real numbers.

\begin{sub}
\label{sub:loading1}
Any high-level span program $\cP$ for $n\times m$-matrices can be implemented as an actual span program with witness size at most $O(\sqrt{nm}\wsize(\cP))$, i.e., the complexity of the subroutine is $O(\sqrt{nm})$. 
\end{sub}

\pfstart
Use $m$ instances $\{\cL_j\}$ of the vector loading subroutine with $e_1,\dots,e_n$ being the standard basis of the vector space of $\cP$. Take the input variables for the subroutine $\cL_j$ from the $j$-th column of $A$.

In the positive case, the contribution of $\cL_j$ is $O(nw_j^2)$, where $w=(w_j)$ is the witness. Hence, the positive witness size of the composed program is $O(n\|w\|^2)$.

In the negative case, with witness $w'$, the contribution of each subroutine is $O(\|w'\|^2)$, because $\{e_i\}$ is a basis. Hence, the negative witness size is $O(m\|w'\|^2)$.

The total witness size can be obtained as the geometrical mean of the maximal positive and negative witness sizes.
\pfend

\subsection{Sparse matrices}
To deal with sparse matrices, we introduce one more subroutine.

\begin{sub}[Demultiplexor]
\label{sub:demux}
There exists a subroutine capable of replacing the specified unit vector of the input space by one of the pivot, namely, the one whose index is specified by the input integer variable. The complete specification of the subroutine is in \reftbl{demux}.
\end{sub}

\begin{table}[htb]
\begin{center}
\begin{tabular}{lp{10cm}}
\hline
Pivots:& vectors $e_0,e_1,\dots, e_{n-1}$.\\
Input variables:& an integer $c$ in range $[0,n-1]$.\\ 
Input space:& one-dimensional, spanned by unit vector $g$.\\
Result:& each input vector $\tilde a_j$ of $\cP_1$ gets transformed into input vector $a_j = \tilde a_j+\langle \tilde a_j,g\rangle (e_c-g)$ of $\cP_2$.\\
Positive witness: & does not change.\\
Positive cost:& $O\left((\log n)\sum_{j=1}^m w_j^2\langle \tilde a_j, g\rangle^2\right)$.\\
Negative witness:& witness $w'$ gets extended to the witness $\tilde w'$ of $\cP_1$ by letting $\langle \tilde w', g\rangle = \langle w', e_c\rangle$.\\
Negative cost:& $O\left((\log n)\sum_{i=1}^n \langle w', e_i\rangle^2\right)$.\\
\hline
\end{tabular}
\caption{Specification of demultiplexor} \label{tbl:demux}
\end{center}
\end{table}

The logarithmic factors in the positive and negative costs arise from encoding $c$ into binary. We believe it is possible to avoid this logarithmic overhead by querying $c$ directly, but we have no means to realize such a query in a span program yet. The subroutine can be used also in the opposite direction, to replace $e_c$ by $g$. Functionally, the subroutine is a multiplexor then. We use it in this mode in \refsub{loading3}. 

\pfstart
For simplicity, assume $n=2^k$, otherwise just truncate the resulting subroutine. Recall that $c$ is given as in~\refeqn{integer}.

For $a=0,1,\dots,k$ and $\ell=0,1,\dots, 2^a-1$, define vector $f^{(a)}_\ell$ as follows. Vector $f^{(0)}_0$ is equal to $g$ and $f^{(k)}_\ell$ to $e_\ell$. For other values of $a$, vectors $f^{(a)}_\ell$ form an orthonormal basis of the working space.

For each $a=0,\dots,k-1$, $b\in\{0,1\}$ and $\ell$ from $\{0,1,\dots,2^a-1\}$, define an input vector 
\begin{equation}
\label{eqn:vectorX}
v_{a,b,\ell}=f^{(a+1)}_{b2^a+\ell} - f^{(a)}_\ell
\end{equation}
that is labeled by value $b$ of variable $c_a$.

In the positive case, it is easy to see that
$$e_c-g = \sum_{a=0}^{k-1} v_{a,c_a,c\bmod 2^a}.$$
Then this vector can be used to replace each $\tilde a_j$ by $\tilde a_j+\langle \tilde a_j,g\rangle (e_c-g)$. This gives the specified positive complexity.

In the negative case, let $w'$ be the witness in $\cP_2$ we extend to the witness $\tilde w'$ of the composed program. All $f^{(a)}_\ell$'s form a full binary tree in a natural way, with $g$ at the root, $e_i$'s at the leaves and with two vertices $f^{(a)}_\ell$ and $f^{(a+1)}_{\ell'}$ connected, iff there is an input vector of~\refeqn{vectorX} being the difference of two. We say an edge is available or false if the corresponding input vector is available or false, respectively.

For each $f^{(a)}_\ell$, there is unique $e_i$ accessible from $f^{(a)}_\ell$ via available edges. Let $\langle \tilde w', f^{(a)}_\ell\rangle$ be equal to $\langle w', e_i\rangle$. In particular, $\langle \tilde w', g\rangle = \langle w', e_c\rangle$. It is easy to see that $\tilde w'$ is orthogonal to all available input vectors of the subroutine. This is because an available edge connects two vertices with the same $e_i$'s chosen.

It remains to estimate the contribution of $\cS$ to the witness size. Each false edge contributes a square of the difference of values of two vertices it connects, i.e., 
$$(\langle w', e_i\rangle - \langle w', e_{i'}\rangle)^2\le 2(\langle w', e_i\rangle^2 + \langle w', e_{i'}\rangle^2)$$
for some $i$ and $i'$. It is easy to see that each vertex is incident to at most 2 false edges and each $e_i$ appears in at most $k+1$ vertices. Hence, the total witness size contribution is at most $(4k+4)\sum_{i=0}^n\langle w',e_i\rangle^2$.
\pfend

By adding this subroutine to \refsub{loading1}, it is easy to get a variant of the latter for sparse input vectors, i.e., matrices with sparse columns.
\begin{sub}
\label{sub:loading2}
Any high-level span program $\cP$ for $n\times m$-matrices can be implemented as an actual span program with witness size at most $O(k\sqrt{m}\wsize(\cP)\log n)$ assuming the following requirements for matrix $A=(a_{ij})$:
\begin{itemize}
\item there are at most $k$ non-zero elements in each column of $A$;
\item the $j$-th column of $A$ is represented by integers $c_{i,j}$ and reals $x_{i,j}$ with $i=1,\dots,k$ so that $a_{c_{i,j},j}=x_{i,j}$ and this specifies all non-zero elements of $A$.
\end{itemize}
\end{sub}

\pfstart
Extend the linear space of $\cP$ with $m$ pairwise orthogonal $k$-dimensional subspaces $\{U_j\}$. Use $m$ vector loading subroutines $\{\cL_j\}$ to load, for each fixed $j$, vectors, given by $x_{i,j}$, into $U_j$. Then use $km$ demultiplexors $\cD_{i,j}$ to replace the $i$-th coordinate of $V_j$ by the $c_{i,j}$-th coordinate of $\cP$. In particular, the pivots of each $\cD_{i,j}$ are the elements of the standard basis of the space of $\cP$.

Consider the positive case first. After composing $\cL_j$, subspace $U_j$ contains a ``compressed version'' of $v_j$. The witness is the same as in $\cP$ as demultiplexors don't affect positive witnesses. Hence, the cost of $\cL_j$ is $O(k w_j^2)$. The cost of $\cD_{i,j}$ is $O(w_j^2 x_{i,j}^2\log n) = O(w_j^2\log n)$, because $x_{i,j}\in[-1,1]$. Thus, the total positive witness size is $O(k \|w\|^2\log n)$.

In the negative case, the cost of each of $\cD_{i,j}$ is $O(\|w'\|^2\log n)$, again because the pivots form a basis. The cost of each $\cL_j$ equals $O(\|\proj_j w'\|^2)$, where $\proj_j$ is the orthogonal projection onto the space spanned by $\{e_{c_{i,j}}\}$ with $i=1,\dots,k$. Clearly, it is at most $O(\|w'\|^2)$. Hence, the total negative witness size is $O(km\|w'\|^2\log n)$. 

Taking the geometric mean, one obtains the total witness size of the actual span program.\pfend

A bit more effort is required to get a matrix loading subroutine for sparse matrices:
\begin{sub}
\label{sub:loading3}
Any high-level span program $\cP$ for $n\times m$-matrices can be implemented as an actual span program with witness size at most $O(\sqrt{k\ell(k+\ell)}\wsize(\cP)\log (m+n))$ assuming the following requirements for matrix $A=(a_{ij})$:
\begin{itemize}
\item there are at most $k$ non-zero elements in each column of $A$;
\item there are at most $\ell$ non-zero elements in each row of $A$;
\item the $j$-th column of $A$ is represented by integers $c_{i,j}$ and reals $x_{i,j}$ with $i=1,\dots,k$ so that $a_{c_{i,j},j}=x_{i,j}$ and this specifies all non-zero elements of $A$ and
\item for each $i=1,\dots,n$ there is a list of integers $d_{i,1},\dots,d_{i,\ell}$ that specifies indices of all non-zero elements in the $i$-th row of $A$. The list may also contain indices of zero elements.
\end{itemize}
\end{sub}

\pfstart
Similarly to the proof of \refsub{loading2}, we define $m$ $k$-dimensional vector subspaces $\{U_j\}$. To that, we add $m$ $n$-dimensional subspaces $\{W_j\}$. Let $f_{i,j}$ and $h_{i,j}$ be the $i$-th elements of the standard bases of $U_j$ and $W_j$, respectively. We compose
\begin{itemize}
\item $m$ vector loading subroutines $\cL_j$ to load vectors specified by $x_{i,j}$'s into $V_j$'s;
\item $km$ demultiplexors $\cD_{i,j}$ replacing $f_{i,j}$ by $h_{c_{i,j},j}$;
\item $n\ell$ demultiplexors $\cM_{i,j}$ replacing $e_i$ by $h_{i,d_{i,j}}.$ Note that these latter demultiplexors work in the direction opposite to the one of \refsub{demux}, so they are, actually, multiplexors.
\end{itemize}

Let us analyse the witness size of this span program. Consider the positive case at first. Various subroutines contribute the following costs towards the total witness size. The $j$-th vector loading subroutine $\cL_j$ contributes $O(kw_j^2)$. Next, by the similar argument as in the proof of \refsub{loading2}, the cost of $\cD_{i,j}$ is $O(w_j^2\log n)$. After composing all $\cL_j$'s and $\cD_{i,j}$'s, each $W_j$ contains input vector $v_j$. Then, they are moved into the vector space of $\cP$, using $\cM_{i,j}$'s. One can estimate the cost of $\cM_{i,j}$'s by noticing that they collect precisely everything put by $\cD_{i,j}$'s into $W_j$'s. Hence, the total cost of all $\cM_{i,j}$'s is $O(k\|w\|^2\log m)$. Summing everything up, we see that the positive witness size of the program is $O(k\|w\|^2\log(m+n))$.

Let us now consider the negative case. Let, as usually, $w'$ be the witness. At first, we will describe how the witness $\tilde w'$ of the composed program looks like, and then calculate the witness size. All non-zero components of $\tilde w'$ are given by: $\langle \tilde w', h_{i,d_{i,j}}\rangle = \langle w', e_i\rangle$ for $j=1,\dots,\ell$ and $\langle \tilde w', f_{i,j}\rangle = \langle w', e_{c_{i,j}}\rangle$ for $i=1,\dots,k$.

Since for each $i$ and $j=1,\dots,\ell$ the condition on the negative witness in \reftbl{demux} for $\cM_{i,j}$ is fulfilled (both sides of the equality equal $\langle w', e_i\rangle$), the witness can be successfully extended to the working space of $\cM_{i,j}$, and its contribution towards the witness size is 
$$O\left((\log m)\sum_{j=1}^m \langle \tilde w', h_{i,j}\rangle^2 \right) = O\left(\ell\langle w', e_i\rangle^2\log m\right).$$
Since there are $l$ multiplexors for each $i$, the total contribution of all $\cM_{i,j}$'s is $O(\ell^2 \|w'\|^2\log m)$. 

Similarly, the cost of $\cD_{i,j}$ equals $O(\log n)$ times the norm squared of the projection of $\tilde w'$ onto $W_j$. Hence, it is not hard to see that the total contribution of all $\cD_{i,j}$'s is $O(k \ell \|w'\|^2 \log n)$. By the same argument, the total contribution of all $\cL_j$'s is $O(\ell \|w'\|^2)$. Hence, finally, the total negative witness size is $O(\ell (k+\ell) \|w'\|^2\log (m+n))$.

One can obtain the witness size of the whole program by taking the geometric mean.
\pfend

\subsection{Lower bounds}
\label{sec:lower}
In this section, we show that $O(\sqrt{n})$ and $O(\sqrt{m})$ factors in Subroutines~\ref{sub:loading1} and~\ref{sub:loading2} are optimal. We do this by converting a high level span program for some problem into a quantum algorithm, using the corresponding matrix loading subroutines. Then we utilize the known lower bound for the quantum algorithm. In fact, we will use one quantum lower bound only: \refthm{grover}.

The first problem is as follows: we are given $m$ $n$-tuples consisting of $\pm 1$'s. Assume $n$ is even, and it is promised that each tuple either contains only 1's, or contains equal amount of 1's and $-1$'s. The question is to detect whether there is a tuple containing only 1's. This is a combination of the Grover and Deutsch-Jozsa~\cite{jozsa} problems.

Clearly, a high level span program with target vector $t=(1,1,\dots,1)$ solves the problem. The positive witness size is at most 1, and the negative witness size is at most $1/n$, as $w'=t/n$ is a negative witness. Hence, the total witness size is $O(1/\sqrt{n})$. But this, as a search problem, requires $\Omega(\sqrt{m})$ quantum queries. Hence, the complexity of the loading subroutine should be at least $\Omega(\sqrt{nm})$ that matches the result of \refsub{loading1}.

One objection to this bound could be that $O(\sqrt{n})$ factor coincides with the norm of vectors being loaded into the span program, and it could be possible that it is needed only for dealing with such large vectors. In fact, it is not so, and we show that this factor is vital even if we require the vectors have norm $O(1)$. We show this in a special case of $m=1$.

Assume the vector space of a high level span program is $\R^{n+1}$, and let $e_0,\dots,e_n$ be an orthonormal basis. Target vector $t$ equals $e_0$. The search for a `1' in a bit string $x=(x_i)$ can be implemented by giving the input vector $e_0+\sum_{i: x_i=1}e_i$ to the span program. If all $x_i$'s are zeroes, the input vector equals the target vector, and the witness size is 1. Otherwise, if one of the input variables, say $x_j$, equals 1, $w' = e_0-e_j$ is a negative witness of size $2$. Hence, the total witness size is $O(1)$. If we require that at most one input variable can be set to 1, the norm of the input vector is bounded by $\sqrt{2}$, but the complexity of loading the input matrix still has to be $\Omega(\sqrt{n})$, in order to match the lower bound for the unique search problem.

What is still unclear, is whether $O(\sqrt{mn})$ bound of \refsub{loading1} is tight if all columns of matrix $A$ have norm $O(1)$ and $m=\omega(1)$. Also, it is an open problem whether dependency on $k$ and $\ell$ in Subroutines~\ref{sub:loading2} and~\ref{sub:loading3} can be improved.

\subsection*{Acknowledgements}
I am grateful to Andris Ambainis for the suggestion to solve the determinant and rank problems with span programs, and for many useful comments during the development of the paper. Also, I would like to thank Mark Meckes for discussion on Wishart matrices.

This work has been supported by the European Social Fund within the project ``Support for Doctoral Studies at University of Latvia''.

\end{document}